\documentclass[aps,showpacs,twocolumn,prb,superscriptaddress]{revtex4-1}
\usepackage[english]{babel}
\usepackage[latin1]{inputenc}
\usepackage{hyperref}
\usepackage{amsmath, amssymb, amsfonts}
\usepackage{amssymb,xcolor}
\usepackage{graphicx}
\usepackage{multirow}
\graphicspath{{images/}}
\usepackage{exscale}
\usepackage{graphicx}
\usepackage{latexsym}
\usepackage{graphicx}  
\usepackage{epstopdf}
\usepackage{float}  
\usepackage{subfigure}  
\usepackage{multirow}
\usepackage{bm}

\newcommand{\expval}[1]{\langle #1 \rangle}

\newcommand{\Tr}{\mathrm{Tr}}

\newcommand{\ii}{\mathrm{i}}

\newcommand{\n}{\nonumber\\}

\newcommand{\dd}{\mathrm{d}}
\begin{document}

\title{Embedded Majorana Islands}

\author{Jin-Xing Hou}
\address{Institute of Modern Physics, Northwest University, Xian 710127, China}
\address{Department of Physics, School of Science, Westlake University, Hangzhou 310030, P. R. China}
\address{Shaanxi Key Laboratory for Theoretical Physics Frontiers, Xian 710127, China}

\author{Alex Weststr\"{o}m}
\address{Department of Physics, School of Science, Westlake University, Hangzhou 310030, P. R. China}
\address{Institute of Natural Sciences, Westlake Institute for Advanced Study, Hangzhou 310024, P. R. China}

\author{Rui Wang}

\address{National Laboratory of Solid State Microstructures and Department of Physics, Nanjing University, Nanjing 210093, China}
\address{Collaborative Innovation Center for Advanced Microstructures, Nanjing 210093, China}

\author{Wen-Li Yang}
\address{Institute of Modern Physics, Northwest University, Xian 710127, China}
\address{Shaanxi Key Laboratory for Theoretical Physics Frontiers, Xian 710127, China}
\address{Peng Huanwu Center for Fundamental Theory, Xian 710127, China}

    \author{Jian Li}
\address{Department of Physics, School of Science, Westlake University, Hangzhou 310030, P. R. China}
\address{Institute of Natural Sciences, Westlake Institute for Advanced Study, Hangzhou 310024, P. R. China}

%


\date{\today}
\begin{abstract}

Mesoscopic superconducting islands hosting Majorana zero modes (MZMs), or Majorana islands in short, offer a prototype of topological qubits. In this work we investigate theoretically the model of a generic Majorana island tunneling-coupled to a single-piece metallic substrate, hence an \textit{embedded Majorana island}. We show the crucial consequences of an interplay between the topological ground states nonlocally addressed by the MZMs and the metallic bath with coherent electron propagation: on the one hand, the topological degeneracy on the Majorana island can be preserved, by virtue of the particle-hole symmetry, despite the apparent bath-induced coupling between MZMs; on the other hand, the electronic interference in the metallic bath may lead to profound alterations to the renormalization group behavior of the hybrid system towards low energy/temperature compared with conventional Kondo physics. This work serves to establish the model of embedded Majorana islands as an experimentally relevant and theoretically intriguing problem particularly in the direction of topological quantum computation.







\end{abstract}

\maketitle

\section{Introduction}\label{set:intro}

Topological quantum computation based on Majorana zero modes (MZMs) remains one of the most challenging but rewarding goals to be achieved in condensed matter physics\cite{kitaev2001unpaired, kitaev2003faulttoleranta, komijani2020isolating, ivanov2001nonabelian}.
In pursuing this goal, several MZM platforms have been established, including semiconductor nanowires \cite{albrecht2016exponentiala,das2012zerobias,lutchyn2018majoranaa,mourik2012signatures,nadj-perge2014observation,prada2020andreev,rokhinson2012fractional,sarma2012splitting,stanescu2011majorana,stanescu2013majoranaa,li_spontaneous_2013,cao2023recent}, magnetic atom lattices \cite{braunecker2013interplay,brydon2015topological,choy2011majorana,heimes2014majorana,klinovaja2013topological,li2014topological,li2016manipulating,mourik2012signaturesa,nadj-perge2013proposal,nadj-perge2014observationa,pawlak2016probing,ruby2015end,vazifeh2013selforganized,zhang2016topological}, superconducting vortices \cite{qin2023two,chan2017generic,chiu2020scalable,flensberg2021engineered,hosur2011majoranaa,kong2020tunable,kong2021emergent,kong2021majorana,li2022controllable,menard2019isolated,mercado2022hightemperature,qin2019topological,roy2010topological,sun2016majoranaa,zhu2020nearly}, Josephson junctions \cite{fu2009josephsona,jiang2011unconventional,kwon2004fractional,liu2021josephson,maiti2015josephson,zhang2016topologicala}, etc.\cite{wilczek2009majorana,fu2008superconducting,beenakker2013search,liu2021minimal}. A major hurdle that plagues all platforms so far, however, is to go beyond the widely-employed signatures, oftentimes as zero-bias 
conductance peaks\cite{law2009majorana}, that are convenient to measure but bear only information relying on the quasi-particle density of states \cite{vayrynen2020signatures,jeon2017distinguishing,li2018majorana,kong2019observation,sun2016majorana} instead of the all-important many-body ground states. To directly demonstrate the presence of well-protected, topologically degenerate ground states in any Majorana platform will be a necessary step as well as an immediate milestone towards realizing a controllable topological qubit, which is the fundamental building block of topological quantum computation.

\begin{figure}[h]
\centering
\includegraphics[width=0.85\linewidth]{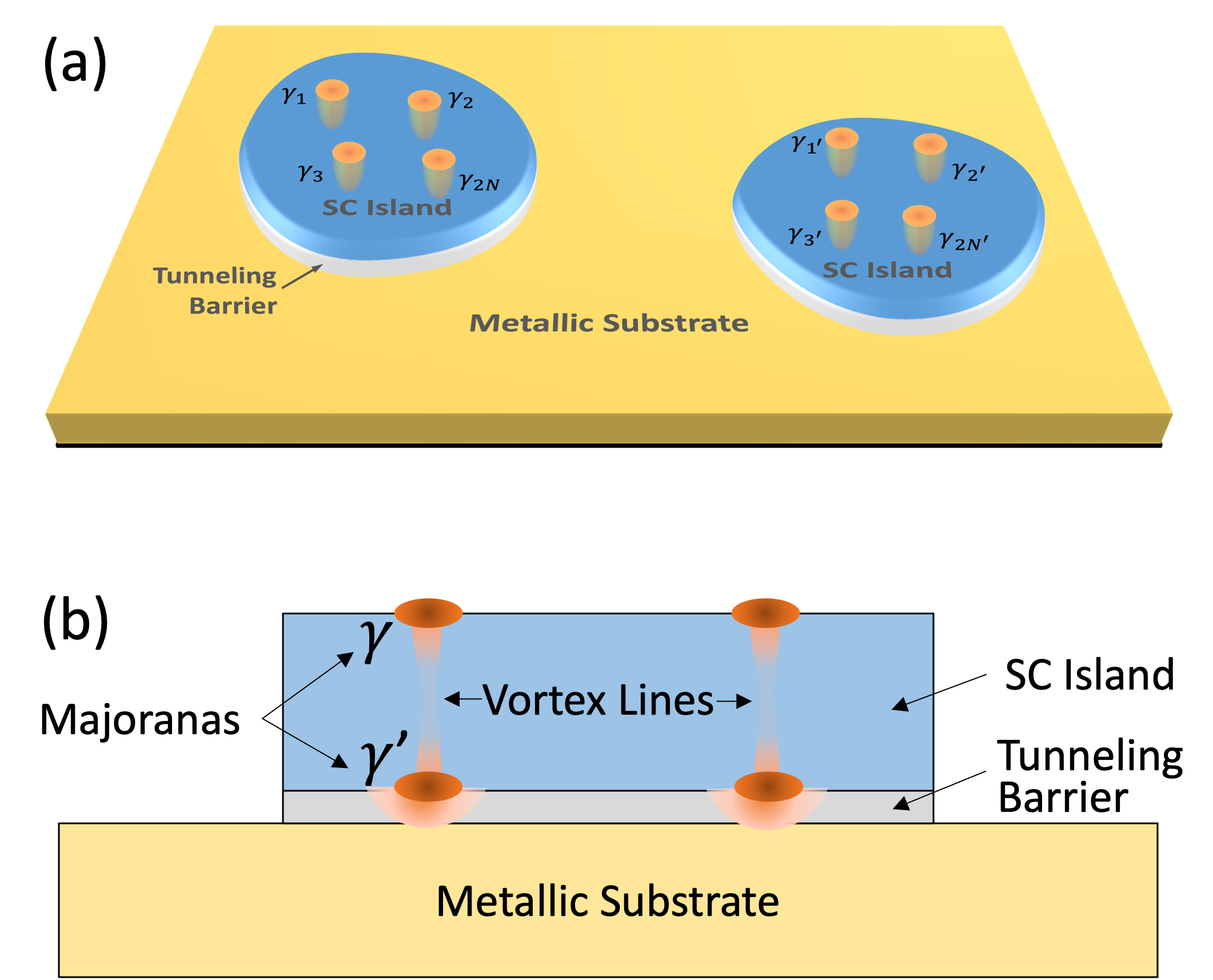}
\caption{(a) Illustration of generic embedded Majorana islands in a metallic substrate with tunneling barriers in-between. Each Majorana island is a superconducting island hosting an even number of Majorana zero modes, a subset of which are in contact with the metallic substrate as an electron bath through quantum tunneling. (b) An example (side view) of an embedded Majorana island with Majorana zero modes induced at the end of the vortex lines in certain iron-based superconductors. Only the bottom Majoranas are coupled to the metallic substrate in this case.}\label{fig:device}
\end{figure}

To that end, one insightful idea pioneered by B\'eri and Cooper \cite{beri2012topological} is to exploit Kondo physics with topological degeneracy. In this setting \cite{beri2012topological,li2023topological,bao2017topologicalb,herviou2016manyterminal,kashuba2015topologicala,tsvelik2014topological,li2023multichannel,papaj2019multichannel,beri2017exact,galpin2014conductance}, a capacitively charged mesoscopic superconducting island that hosts a certain number of Majorana zero modes, or a Majorana qubit by general definition, acts like a Kondo impurity when tunneling-coupled to electron reservoirs via metallic leads and exhibits prototypical Kondo features such as non-Fermi liquid temperature dependence of conductance \cite{beri2012topological,bao2017topologicalb,beri2017exactc,galpin2014conductancea,kashuba2015topologicala,li2023multichannel,li2023topological,snizhko2018parafermionica,tsvelik2014topological}. Restrictions exist, however, in current research along this line with metallic leads being essentially one-dimensional and addressing MZMs individually. While such a configuration is naturally implemented in some Majorana platforms such as the nanowires, it is hardly feasible for others like those based on superconducting vortices. In the latter type of Majorana platforms, superconducting islands are typically formed on top of larger substrates, giving rise to a scenario of \textit{embedded Majorana islands} (see Fig.~\ref{fig:device}) when the substrate is composed of a thin insulating layer, serving as a tunneling barrier, and an extended metallic layer, serving as a higher-dimensional, simply-connected electron bath. More importantly, besides its feasibility, this scenario permits quantum interference effect in the metallic substrate that can significantly enrich the physics by taking into account the nonlocal nature of the Majorana qubit states and can be readily probed by established transport measurement techniques or scanning tunneling microscopes.

In this work, we investigate the scenario of {embedded Majorana islands} from three basic aspects: the validity and robustness of the topological degeneracy, the effective Hamiltonian and its mapping to Kondo-type models, and the renormalization group (RG) flow of the island-bath coupling term. After introducing in Sec.~\ref{sec:model} the model Hamiltonian for a Majorana island tunneling embedded in a metallic bath, we show in Sec.~\ref{sec:topo_deg} that, rather surprisingly, topological degeneracy in a Majorana island is preserved, albeit broadened, when the coupling effect between MZMs and the metallic bath is included --- this justifies the starting point of our work. In Sec.~\ref{sec:ham_eff}, we establish the algebraic conditions for our model to be mapped to an $\mathfrak{so}(M)$ Kondo model. The mapped model is not necessarily isotropic, and can be subject to a parity constraint which further leads to overscreening of the topologically degenerate ground states by the bath electrons. In Sec.~\ref{sec:rg}, we proceed to derive the effective coupling Hamiltonian at the low-energy limit by performing a perturbative RG analysis up to the third order corrections. Here, the effect of quantum interference in the metallic substrate becomes prominent and leads to significantly different physical consequence in the RG flow compared with previously known results. In Sec.~\ref{sec:discussion} we discuss the connection between the various aspects investigated in this work as well as their limitations, and identify questions to be further addressed for a more complete understanding and characterization of embedded Majorana islands.

\section{The model of embedded Majorana island}\label{sec:model}

Throughout this paper, we consider generic formations of superconducting islands, where MZMs can be induced by one of the currently available schemes (e.g. vortices in Fe-based superconductors\cite{hirschfeld2011gap, kong2021emergent, li2022exploring, liu2023roadmap, sang2022majorana, song2022phasemanipulationinduced, wang2018evidence, wang2020evidence}), on a metallic substrate with a tunneling barrier layer in-between (see Fig.~\ref{fig:device}). We assume that a subset, with number $M$, of the totally $M_0$ MZMs on each island (assumed to be independent) are tunneling coupled to the single-piece metallic substrate. The fact that the metallic substrate, or metallic bath, has a simply-connected topology distinguishes our setup, and hence its consequential physics, from the previously investigated ones where the metallic baths are realized by separate, disconnected metallic leads \cite{beri2012topological,li2023topological,bao2017topologicalb,herviou2016manyterminal,kashuba2015topologicala,tsvelik2014topological,li2023multichannel,papaj2019multichannel,beri2017exact,galpin2014conductance}. The advantage of our setup is two-fold: the first is its practical relevance for Majorana platforms beyond the nanowire systems, particularly when additional leads coupled to individual MZMs are virtually impossible such as for the vortex systems; the second is that our setup allows for a significantly extended set of experimental tools and signatures, including bulk thermodynamic properties and scanning tunneling spectroscopy, for the measurement of topological features associated with Majorana islands, or the detection of Majorana qubits eventually.

Specifically, the model Hamiltonian of our setup consists of four parts:
\begin{equation}\label{ham_ful}
  H =  H_{I} + H_{C} + H_{B} + H_{T},
\end{equation}
where
\begin{equation}\label{eq:HM}
  H_I = \ii\sum_{\alpha,\beta=1}^{M_0}\xi_{\alpha\beta}\gamma_\alpha\gamma_\beta
\end{equation}
is the low-energy effective Hamiltonian for an isolated Majorana island where pairs of MZMs (labeled by $\alpha$, $\beta$) are possibly coupled with strength $\xi_{\alpha\beta}$ --- when the MZMs are sufficiently separated, $\xi_{\alpha\beta}$ is exponentially suppressed and therefore will be neglected hereafter;
\begin{equation}\label{eq:HC}
  H_C = E_C(\hat{N}-N_0)^2
\end{equation}
represents the total charging energy for a Majorana island induced by additional gating (possibly by taking advantage of the metallic substrate) with $E_C$ the charging energy, $\hat{N}$ the island electron number operator and $N_0$ a number controlled by the gating;
\begin{equation}\label{eq:HB}
  H_{B}=\sum_{\bm{k}} \xi_{\bm{k}} c_{\bm{k}}^\dag c_{\bm{k}},
\end{equation}
with $\xi_{\bm{k}} = {k^2}/{2m}-\mu$ is the Hamiltonian for the metallic bath, assumed to be an ideal metal ignoring spin because currently all experimentally realized Majorana platforms have broken time-reversal symmetry;
\begin{equation}\label{eq:HT0}
  H_T = \sum_{\alpha=1}^{M\le M_0} \int \dd \bm{r} \lambda_{\alpha}(\bm{r})\mathrm{e}^{\mathrm{i}\phi/2}\gamma_\alpha c_{\bm{r}}+\mathrm{h.c.}
\end{equation}
is the local tunneling term between the Majorana island and the metallic bath with the tunneling rates given by real, non-negative functions $\lambda_{\alpha}(\bm{r})$ and the tunneling phase factor $\mathrm{e}^{\mathrm{i}\phi/2}$ tied to the superconducting condensate\cite{fu2010electron}. 

In this work, we assume $\mu \gg \Delta, E_C \gg \lambda_\alpha, T \gg \xi_{\alpha\beta}$, where $\Delta$ is the superconducting gap or more generally the excitation gap between the MZMs and any nontopological finite energy quasiparticles in the superconductor, $T$ is the temperature. Aligned with the assumption that $\xi_{\alpha\beta}$ is negligible, here we assume that the MZMs are separated far enough such that the overlap between $\lambda_{\alpha}$ and $\lambda_{\beta}$ for $\alpha\ne\beta$ is also negligible.

\section{Topological degeneracy in the presence of metallic bath}\label{sec:topo_deg}

One most prominent effect of the simply-connected metallic bath in our setup is its capability to mediate interaction between MZMs analogous to the Ruderman-Kittel-Kasuya-Yosida (RKKY) mechanism for the interaction of local magnetic moments mediated by itinerant electrons. This, a priori, can lead to lifting of topological degeneracy in the Majorana island even if the degeneracy is well maintained by keeping the inherent MZM coupling strength $\xi_{\alpha\beta}$ small compared with all other energy scales. In this section, we address this potential problem and show that, by virtue of the particle-hole symmetry, the topological degeneracy remains valid despite the coupling between the Majorana island and the bath.

For the coupled system here, it is well-known that the effect of the bath, which is assumed to be an ideal, non-interacting metallic reservoir, can be theoretically lumped into a self-energy term on top of the interacting, impurity-like subsystem which is the Majorana island in our case. Such a self-energy only depends on $H_B$ and $H_T$ in Eqs.~\eqref{eq:HB} and \eqref{eq:HT0}, and can be expressed as
\begin{align}\label{eq:self_energy}
  \Sigma_{\alpha \beta}(\omega) &= \int \dd\bm{r} \dd \bm{r}^\prime\lambda_\alpha(\bm{r}) \lambda_{\beta}(\bm{r}^\prime) \nonumber\\
  &\qquad\times\left[ G_{0} (c_{\bm{r}},c_{\bm{r}'}^\dag;\omega^+) + G_{0} (c_{\bm{r}}^\dag,c_{\bm{r}'};\omega^+)\right],
\end{align}
where $G_{0}(A,B;\omega^+\equiv\omega+\ii 0_+)$ is the retarded correlation function between $A$ and $B$ in the pristine metal. The two correlation functions occurring in the integrand of the above expression are related by the particle-hole symmetry (see Appendix \ref{app:phs} for the derivation)
\begin{align}\label{eq:p_h_symmetry}
  G_{0} (c_{\bm{r}}^\dag,c_{\bm{r}'};\omega^+) = - G_{0} (c_{\bm{r}},c_{\bm{r}'}^\dag;(-\omega)^+)^*.
\end{align}
Furthermore, the assumption of an ideal metal implies that $G_{0}(c_{\bm{r}},c_{\bm{r}'}^\dag;\omega^+)$ can be effectively simplified to $G_{0}(|\bm{r}-\bm{r}'|)$. Namely, $G_{0}$ is taken to be isotropic and approximated by its Fermi surface values ($\omega\rightarrow 0$) within the energy scale considered in the present work. In the case of an infinite, three-dimensional ideal metal, for example, $G_{0}(|\bm{r}-\bm{r}'|) = -\pi\rho_0 \mathrm{e}^{\ii k_F |\bm{r}-\bm{r}'|}/(k_F |\bm{r}-\bm{r}'|)$ with $\rho_0$ the density of states at the Fermi energy and $k_F$ the Fermi wave-vector. It follows that the self-energy is in practice energy-independent and, more importantly, anti-hermitian (both purely imaginary and symmetric):
\begin{align}
  &\Sigma_{\alpha \beta} = \Sigma_{\beta \alpha} = -\Sigma_{\alpha \beta}^* \nonumber\\
  &\;= 2\ii\int \dd\bm{r} \dd \bm{r}^\prime\lambda_\alpha(\bm{r}) \lambda_{\beta}(\bm{r}^\prime)
  \Im G_{0}(|\bm{r}-\bm{r}'|).
\end{align}
This further implies that the self-energy term can be diagonalized by an orthogonal matrix (namely, by recombination of Majorana operators with real coefficients) with purely imaginary eigenvalues. In other words, the coupling between the metallic substrate and the Majorana island will not lift the topological degeneracy but only broaden it with an energy scale proportional to $\rho_0\lambda^2$.

\section{Effective Coupling Hamiltonian}\label{sec:ham_eff}

In this section, we derive the effective coupling Hamiltonian between the Majorana island and the metallic bath in its most general form and discuss its mapping to Kondo-type coupling terms.
For simplicity, we now assume
\begin{align}
    \lambda_{\alpha} (\bm{r}) = \lambda_\alpha \delta(\bm{r}-\bm{r}_\alpha), \quad (\lambda_\alpha\ge 0)
\end{align}
which will not alter the physics discussed in the later part of this paper as long as the overlap between the original functions $\lambda_{\alpha}(\bm{r})$ and $\lambda_{\beta}(\bm{r})$ is negligible for different $\alpha$ and $\beta$. This immediately implies that the local tunnelling term Eq.~\eqref{eq:HT0} simplifies to
\begin{equation}\label{eq:HT}
  H_T = \sum_{\alpha=1}^{M} \lambda_{\alpha}\mathrm{e}^{\mathrm{i}\phi/2}\gamma_\alpha c_{\bm{r}_\alpha}+\mathrm{h.c.}.
\end{equation}
Consequently, the effective coupling between the Majorana island and the metallic bath, when projected to the ground state subspace with only $N_0$ electrons on the Majorana island but taking into account the virtual processes involving the first excited states with $N_0\pm 1$ electrons, can be obtained by the standard Schrieffer-Wolff transformation\cite{schrieffer1966relation} to be
\begin{align}\label{eq:HIB0}
    H_{IB}^{(0)} = \sum_{\alpha\ne\beta}\Lambda_{\alpha\beta} \gamma_\alpha\gamma_\beta c_{\bm{r}_{\beta}}^\dag c_{\bm{r}_{\alpha}},
\end{align}
where $\Lambda_{\alpha\beta} = {2\lambda_\alpha\lambda_\beta}/{E_c}$ is the bare effective island-bath coupling coefficient not yet including the higher-order processes in the metallic bath. Here and afterwards, the Majorana indices ($\alpha$, $\beta$ etc.) are summed implicitly from 1 to $M$ unless specified otherwise.

Whereas the above form of effective coupling is sufficient for disconnected metallic baths that are in contact with each of the Majoranas individually, as most of the previous studies assume\cite{beri2012topological}, it is no longer so in a low-energy effective theory as soon as the bath becomes connected and electrons can propagate between contact points (say, $\bm{r}_{\beta}$ and $\bm{r}_{\alpha}$) \textit{within} the metal. With a detailed discussion of this key issue postponed to Sec.~\ref{sec:rg} where a systematic renormalization group analysis is performed, here we straightforwardly generalize Eq.~\eqref{eq:HIB0} to the following form:
\begin{align}\label{eq:HIB}
    H_{IB} = \sum_{\alpha\ne\beta\atop\mu\ne\nu}A_{\alpha\beta}^{\mu\nu} \gamma_\alpha\gamma_\beta c_{\bm{r}_{\nu}}^\dag c_{\bm{r}_{\mu}},
\end{align}
where $A_{\alpha\beta}^{\mu\nu}$ is a rank-4 tensor satisfying the symmetry constraint $A_{\alpha\beta}^{\mu\nu} = -A_{\beta\alpha}^{\mu\nu} = A_{\beta\alpha}^{\nu\mu}$. It is easy to check that when  $A_{\alpha\beta}^{\mu\nu} = \frac{1}{2}\Lambda_{\alpha\beta}(\delta_{\alpha}^{\mu}\delta_{\beta}^{\nu}-\delta_{\alpha}^{\nu}\delta_{\beta}^{\mu})$, Eq.~\eqref{eq:HIB} reduces to Eq.~\eqref{eq:HIB0}. In the following we proceed to discuss the mapping of the generalized island-bath coupling to one of the Kondo type. Such a mapping is plausible for at least two reasons: first, as showcased by the topological Kondo effect\cite{beri2012topological}, the physics of the coupling between topologically degenerate ground states and metallic bath is intimately related to generalized Kondo physics; second, the mapping to well-studied Kondo-type models will allow the known experimental signatures in Kondo systems to be transferred to the experimental investigations of the present system.

An obvious starting point to establish the mapping is to notice that the paired Majorana operators $\Gamma_{\alpha\beta} = \ii\gamma_{\alpha}\gamma_{\beta}/2$ ($\alpha,\beta = 1,...,M$) fulfill an $\mathfrak{so}(M)$ algebra\cite{wilczek2009majorana}
\begin{align}
    [\Gamma_{\alpha\beta},\; \Gamma_{\alpha'\beta'}] = \ii \sum_{\alpha''\beta''}C_{\alpha\beta,\alpha'\beta'}^{\alpha''\beta''}\Gamma_{\alpha''\beta''},
\end{align}
where $C_{\alpha\beta,\alpha'\beta'}^{\alpha''\beta''} = \delta_{[\alpha}^{\alpha''}\delta_{\beta][\alpha'}\delta_{\beta']}^{\beta''}$, with $[\cdot\cdot]$ standing for anti-symmetrized indices, is the structural constant for the $\mathfrak{so}(M)$ algebra. Note that the above equation is invariant under an orthogonal transformation $\tilde{\gamma}_\alpha = \sum_{\alpha'}{\gamma}_{\alpha'}R_{\alpha'\alpha}$. It follows that a generalized, but not necessarily isotropic, Kondo form of the generic island-bath coupling term in Eq.~\eqref{eq:HIB}, which reads
\begin{align}\label{eq:HIBK}
    H_{K} &= \sum_{\alpha\ne\beta}\mathcal{J}_{\alpha\beta}\Gamma_{\alpha\beta} \mathcal{S}_{\alpha\beta}, \\
    \mathcal{S}_{\alpha\beta} &= -2\ii\sum_{\mu\ne\nu} ({A}_{\alpha\beta}^{\mu\nu}/\mathcal{J}_{\alpha\beta}) c_{\bm{r}_{\nu}}^\dag c_{\bm{r}_{\mu}},
\end{align}
with the coupling coefficients $\mathcal{J}_{\alpha\beta} = \mathcal{J}_{\beta\alpha}$ to be determined, can be obtained by requiring the bath components to fulfill the same algebra
\begin{align}
    [\mathcal{S}_{\alpha\beta},\; \mathcal{S}_{\alpha'\beta'}] = \ii \sum_{\alpha''\beta''}C_{\alpha\beta,\alpha'\beta'}^{\alpha''\beta''}\mathcal{S}_{\alpha''\beta''}.
\end{align}
This in turn requires
\begin{align}\label{eq:Acom}
    [\bm{A}_{\alpha\beta},\; \bm{A}_{\alpha'\beta'}] = \sum_{\alpha''\beta''}C_{\alpha\beta,\alpha'\beta'}^{\alpha''\beta''}\frac{\mathcal{J}_{\alpha\beta}\mathcal{J}_{\alpha'\beta'}}{2\mathcal{J}_{\alpha''\beta''}}\bm{A}_{\alpha''\beta''},
\end{align}
where $\bm{A}_{\alpha\beta}$ shall be understood as a matrix labeled by $(\alpha\beta)$ with its matrix element given by ${A}_{\alpha\beta}^{\mu\nu}$. It is easy to check that if ${A}$ is diagonal, namely if ${A}_{\alpha\beta}^{\mu\nu} = (\mathcal{J}_{\alpha\beta}/2){\delta}_{\alpha\beta}^{\mu\nu}$, the above equation is satisfied.

The coupling form in Eq.~\eqref{eq:HIBK} realizes an $\mathfrak{so}(M)$ Kondo model if only the $\Gamma$'s are free, which is indeed the case if $M<M_0$. If $M=M_0$, however, $\Gamma$'s are subject to the parity constraint in our model
\begin{align}\label{eq:parity}
    \ii^{M_0/2}\prod_\alpha \gamma_\alpha = (-1)^{N_0}.
\end{align}
The consequence of this constraint is best illustrated when $M=M_0=4$, where, assuming $N_0$ to be even and without loss of generality, we can identify $\Gamma_{23} = \Gamma_{14} = \sigma_1/2$, $\Gamma_{31} = \Gamma_{24} = \sigma_2/2$ and $\Gamma_{21} = \Gamma_{43} = \sigma_3/2$ with $\sigma_{1,2,3}$ being Pauli operators acting on the fixed-parity two-fold degenerate ground states of the Majorana island. The coupling term in Eq.~\eqref{eq:HIBK} then becomes $H_K = \sigma_1(\mathcal{J}_{23}\mathcal{S}_{23}+\mathcal{J}_{14}\mathcal{S}_{14})+\sigma_2(\mathcal{J}_{31}\mathcal{S}_{31}+\mathcal{J}_{24}\mathcal{S}_{24})+\sigma_3(\mathcal{J}_{21}\mathcal{S}_{21}+\mathcal{J}_{43}\mathcal{S}_{43})$. At this point it is helpful to invoke the well-known fact that the $\mathfrak{so}(4)$ algebra contains exactly two $\mathfrak{su}(2)$ algebras. In the context of $\{\mathcal{S}_{\alpha\beta}\}$, this can be seen as either subset $\{\mathcal{S}_{i=1,2,3}^{+}\}$ or $\{\mathcal{S}_{i=1,2,3}^{-}\}$, where $\mathcal{S}_{1}^{\pm}=(\mathcal{S}_{23}\pm\mathcal{S}_{14})/2$, $\mathcal{S}_{2}^{\pm}=(\mathcal{S}_{31}\pm\mathcal{S}_{24})/2$ and $\mathcal{S}_{3}^{\pm}=(\mathcal{S}_{21}\pm\mathcal{S}_{43})/2$, generates an individual $\mathfrak{su}(2)$ algebra with $[\mathcal{S}_{i}^{\pm}, \mathcal{S}_{j}^{\pm}] = \ii\varepsilon_{ijk}\mathcal{S}_{k}^{\pm}$ and $[\mathcal{S}_{i}^{+}, \mathcal{S}_{j}^{-}] = 0$. It follows that the coupling term can be further rewritten as
\begin{align}\label{eq:HK4}
    H_{K}\Bigr|_{M=M_0=4} &= \sum_{i=1,2,3}\sigma_i \left[\mathcal{J}_{i}^{+}\mathcal{S}_{i}^{+}+\mathcal{J}_{i}^{-}\mathcal{S}_{i}^{-}\right],
\end{align}
where $\mathcal{J}_{1}^{\pm}=\mathcal{J}_{23}\pm\mathcal{J}_{14}$, $\mathcal{J}_{2}^{\pm}=\mathcal{J}_{31}\pm\mathcal{J}_{24}$ and $\mathcal{J}_{3}^{\pm}=\mathcal{J}_{21}\pm\mathcal{J}_{43}$.
The physics of the above Hamiltonian is made more transparent by assuming ${A}_{\alpha\beta}^{\mu\nu} = (\mathcal{J}_{\alpha\beta}/2){\delta}_{\alpha\beta}^{\mu\nu}$, such that $\mathcal{S}_{\alpha\beta} = \ii(c_{\bm{r}_{\alpha}}^\dag c_{\bm{r}_{\beta}}-c_{\bm{r}_{\beta}}^\dag c_{\bm{r}_{\alpha}})$, and performing a unitary transformation
\begin{align}
    \tilde{\bm{c}} = \frac{1}{\sqrt{2}}
    \begin{pmatrix}
        1 & -\ii & 0 & 0 \\
        0 & 0 & 1 & -\ii \\
        0 & 0 & 1 & \ii \\
        -1 & -\ii & 0 & 0
    \end{pmatrix}
    \begin{pmatrix}
        {c}_{\bm{r}_{1}} \\
        {c}_{\bm{r}_{2}} \\
        {c}_{\bm{r}_{3}} \\
        {c}_{\bm{r}_{4}}
    \end{pmatrix}.
\end{align}
We obtain for $M=M_0=4$:
\begin{align}\label{eq:HIBK4}
    H_{K}
    &= \sum_{i=1,2,3}\sigma_i \tilde{\bm{c}}^\dag\left[\mathcal{J}_{i}^{+}{s}_{i}^{+}\otimes{s}_{0}^{-}+\mathcal{J}_{i}^{-}{s}_{0}^{+}\otimes{s}_{i}^{-}\right]\tilde{\bm{c}},
\end{align}
where ${s}_{0,1,2,3}^{\pm}$ are Pauli matrices (including identity) acting on the ``$+/-$'' copies of the pseudo-spin-$1/2$ degrees of freedom, respectively. It is clear that these two copies of pseudo-spin-$1/2$ form two competing symmetric, two-channel Kondo coupling. In the case of isotropic coupling $\mathcal{J}_{\alpha\ne\beta}=\mathcal{J}_0$, such that $\mathcal{J}_{i}^{-}$ vanishes identically, only the ``$+$'' copy will remain and the system then reduces to a conventional symmetric two-channel Kondo model \cite{nozieres1980kondo,affleck1993exact}. By contrast, suppose $M=3<M_0$, the parity constraint Eq.~\eqref{eq:parity} is ineffective in reducing the island degrees of freedom addressed by $\{\Gamma_{\alpha\beta}\}$ with $\alpha,\beta=1,2,3$ and $\alpha\ne\beta$, and the system exhibits a single-channel Kondo physics as discovered by B\'eri and Cooper\cite{beri2012topological}.

\section{Scaling analysis}\label{sec:rg}

\begin{figure}[t]
\centering
\subfigure[]{
\begin{minipage}[t]{1\linewidth}
\centering
\includegraphics[width=0.9\textwidth]{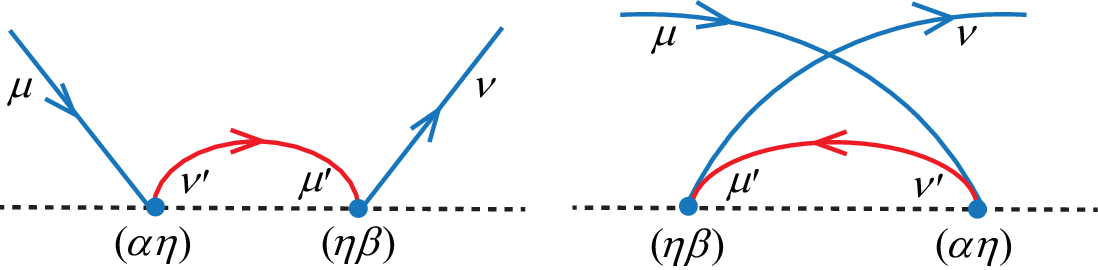}
\label{fig:RG_diagram_2_order}
\end{minipage}%
}%

\subfigure[]{
\begin{minipage}[t]{1\linewidth}
\centering
\includegraphics[width=0.9\textwidth]{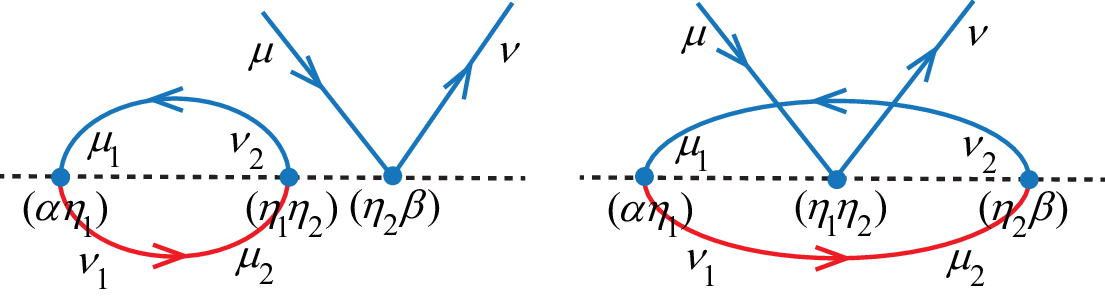}
\label{fig:RG_diagram_3_order}
\end{minipage}%
}%
\centering
\caption{Feynman diagrams that represent the relevant scattering processes of (a) the second and (b) the third orders contributing to the renormalization of the island-bath coupling coefficients at increasing lower energy. The solid and the dashed lines denote the free propagators of the bath electrons and the Majorana island ground states, respectively. The propagators in red belong to the high-energy shell to be integrated out. Each vertex in the diagrams has been labeled by the four indices associated with the coupling coefficient.}
\label{fig:RG_diagram}
\end{figure}

As precedently advertised, the low-energy behavior of the island-bath coupling can be strongly renormalized in the presence of the connected metallic bath. To illustrate this point, let us consider a simplest second-order process involving two island-bath vertices described by the first-order coupling Hamiltonian Eq.~\eqref{eq:HIB0}, which overall amounts to a consecutive action of $\gamma_\alpha\gamma_\beta$ and $\gamma_{\alpha'}\gamma_{\beta'}$ on the island state as well as the scattering of a bath electron from $\bm{r}_{\alpha}$ eventually to $\bm{r}_{\beta'}$ with the total scattering amplitude being $\Lambda_{\alpha'\beta'}\Lambda_{\alpha\beta}\expval{c_{\bm{r}_{\alpha'}}c^\dag_{\bm{r}_{\beta}}}$, where $\expval{c_{\bm{r}_{\alpha'}}c^\dag_{\bm{r}_{\beta}}}$ is the intermediate electron propagator from $\bm{r}_{\beta}$ to $\bm{r}_{\alpha'}$. Particularly, let us consider a special case when $\alpha=\alpha'$, $\beta\ne\beta'$, such that this specific second-order process can be expressed as $\Lambda_{\alpha\beta'}\Lambda_{\alpha\beta}\expval{c_{\bm{r}_{\alpha}}c^\dag_{\bm{r}_{\beta}}}\gamma_{\beta}\gamma_{\beta'}c^\dag_{\bm{r}_{\beta'}}c_{\bm{r}_{\alpha}}$. Compared with the original form in Eq.~\eqref{eq:HIB0}, this clearly induces off-diagonal scattering channels in the sense that the pair of indices $(\beta,\beta')$ for the Majorana island operators is not equivalent to the pair of indices $(\alpha,\beta')$ for the metallic bath operators.

The above heuristic argument can be made precise by performing a systematic scaling analysis to obtain the actual low-energy effective form of the island-bath coupling in Eq.~\eqref{eq:HIB}. Our analysis follows the standard renormalization group approach by integrating out the high-energy electronic states in the metallic bath up to the third order corrections to the general coupling Hamiltonian in Eq.~\eqref{eq:HIB} (see Fig.~\ref{fig:RG_diagram}, and Appendix \ref{app:rgeq} for more details). This leads to the scaling equation for the island-bath coupling coefficients $A_{\alpha\beta}^{\mu\nu}$, given by
\begin{align}\label{eq:RG_org}
  \frac{\dd A_{\alpha\beta}^{\mu\nu}}{\dd l} &= 2(P_{\alpha\beta}^{\mu\nu} - Q_{\alpha\beta}^{\mu\nu}) -(\alpha\leftrightarrow\beta), \\
  P_{\alpha\beta}^{\mu\nu} &= \sum_{\eta\mu'\nu'}f_{\mu'\nu'}A_{\alpha\eta}^{\mu\nu'}A_{\eta\beta}^{\mu'\nu}, \\
  Q_{\alpha\beta}^{\mu\nu} &= \sum_{\eta_{1}\eta_{2} \mu_{1}\mu_{2}\nu_{1}\nu_{2}} f_{\mu_1\nu_2}f_{\mu_2\nu_1} \nonumber\\
  &\qquad\quad \times A_{\alpha\eta_1}^{\mu_1\nu_1}(A_{\eta_1\eta_2}^{\mu_2\nu_2} A_{\eta_2\beta}^{\mu\nu}
  - A_{\eta_1\eta_2}^{\mu\nu} A_{\eta_2\beta}^{\mu_2\nu_2}), \\
  f_{\mu\nu} &= \frac{1}{\rho_0}\int \dd\bm{p}\, \mathrm{e}^{\ii \bm{p}\cdot (\bm{r}_{\mu}-\bm{r}_{\nu})} \delta(\xi_{\bm{p}}-\varepsilon_F),
\end{align}
where $l=-\ln{(D/D_0)}$ is the scaling parameter, and we have absorbed the bath electron density of states $\rho_0$ (assumed to be a constant) into $A$ such that $A$ becomes dimensionless. Here, $f_{\mu\nu}=f_{\nu\mu}$ are symmetric form factors that capture the important electronic interference effect in the metallic bath. $f$ as a function of $\delta r = |\bm{r}_{\mu}-\bm{r}_{\nu}|$ involves an integration over the Fermi surface of the metal and decays rapidly as $\delta r$ increases. The initial condition of the above scaling equation is given by Eq.~\eqref{eq:HIB0}:
\begin{align}
    A_{\alpha\beta}^{\mu\nu}(l=0) = \frac{1}{2}\Lambda_{\alpha\beta}\delta_{\alpha\beta}^{\mu\nu} = \frac{\lambda_\alpha\lambda_\beta}{E_c}\delta_{\alpha\beta}^{\mu\nu},
\end{align}
where $\delta_{\alpha\beta}^{\mu\nu}\equiv \delta_{\alpha}^{\mu}\delta_{\beta}^{\nu}-\delta_{\alpha}^{\nu}\delta_{\beta}^{\mu}$ is the generalized Kronecker delta.

The scaling equation \eqref{eq:RG_org} is difficult to solve analytically in general when the off-diagonal matrix elements of $f$ are finite, as implied by the connectedness of the metallic bath. Progress can be made, however, by the following transformation
\begin{align}\label{eq:tfm_A2A_tilde}
    A_{\alpha\beta}^{\mu\nu} = \sum_{\alpha'\beta'\mu'\nu'}R_{\alpha\alpha'}R_{\beta\beta'}S_{\mu\mu'}S_{\nu\nu'}\tilde{A}_{\alpha'\beta'}^{\mu'\nu'},
\end{align}
where $R$ is real orthogonal and diagonalizes the symmetric matrix of elements $\lambda_{\alpha}\lambda_{\beta}f_{\alpha\beta}$ (which is positive definite as long as the variance of $\lambda$'s is small compared with their average and the off-diagonal entries of $f$ are small compared with its diagonal entries) such that
\begin{align}
    \sum_{\alpha'\beta'}(\lambda_{\alpha'}\lambda_{\beta'}f_{\alpha'\beta'})R_{\alpha'\alpha}R_{\beta'\beta} = g_{\alpha}^2\delta_{\alpha\beta},
\end{align}
and $S_{\mu\mu'} = \lambda_{\mu}R_{\mu\mu'}g_{\mu'}^{-1}$. Note that $S$ is in general a real but non-orthogonal matrix although $R$ is orthogonal. By using $\sum_{\eta}R_{\eta\alpha}R_{\eta\beta}=\delta_{\alpha\beta}$ and $\sum_{\mu'\nu'}f_{\mu'\nu'}S_{\mu'\mu}S_{\nu'\nu}=\delta_{\mu\nu}$, it is straightforward to show that this transformation removes the explicit dependence of $f$ in the scaling equation for $\tilde{A}$, namely,
\begin{align}\label{eq:RG_tfm}
  \frac{\dd \tilde{A}_{\alpha\beta}^{\mu\nu}}{\dd l} &= 2(\tilde{P}_{\alpha\beta}^{\mu\nu} - \tilde{Q}_{\alpha\beta}^{\mu\nu}) -(\alpha\leftrightarrow\beta), \\
  \tilde{P}_{\alpha\beta}^{\mu\nu} &= \sum_{\eta\eta'}\tilde{A}_{\alpha\eta}^{\mu\eta'}\tilde{A}_{\eta\beta}^{\eta'\nu}, \\
  \tilde{Q}_{\alpha\beta}^{\mu\nu} &= \sum_{\eta_{1}\eta_{2} \mu_{1}\nu_{1}} \tilde{A}_{\alpha\eta_1}^{\mu_1\nu_1}(\tilde{A}_{\eta_1\eta_2}^{\nu_1\mu_1} \tilde{A}_{\eta_2\beta}^{\mu\nu}
  - \tilde{A}_{\eta_1\eta_2}^{\mu\nu} \tilde{A}_{\eta_2\beta}^{\nu_1\mu_1}).
\end{align}
More importantly, the initial condition for $\tilde{A}_{\alpha\beta}^{\mu\nu}$ is still diagonal
\begin{align}
    \tilde{A}_{\alpha\beta}^{\mu\nu}(l=0) = \frac{g_\alpha g_\beta}{E_c}\delta_{\alpha\beta}^{\mu\nu},
\end{align}
and this diagonal form is preserved by the RG flow of Eq.~\eqref{eq:RG_tfm}. That is, $\tilde{A}_{\alpha\beta}^{\mu\nu}(l) = J_{\alpha\beta}(l)\delta_{\alpha\beta}^{\mu\nu}$ for all $l$ where $J_{\alpha\beta}$ is a real symmetric matrix depending on $l$. It follows that the scaling equation can be further reduced to a much simpler form in terms of $J_{\alpha\ne\beta}$ (with $J_{\alpha=\beta}=0$ identically):
\begin{align}\label{eq:RG_J}
    &\frac{\dd J_{\alpha\beta}}{\dd l} = (J^2)_{\alpha\beta} + 4(J_{\alpha\beta})^3 - 2J_{\alpha\beta}[(J^2)_{\alpha\alpha}+(J^2)_{\beta\beta}], \\
    &J_{\alpha\beta}(l=0) = \frac{g_\alpha g_\beta}{E_c}, \qquad(\alpha\ne\beta).
\end{align}
Note that similar to Eq.~\eqref{eq:RG_org}, we have absorbed the bath electron density of states $\rho_0$ (assumed to be a constant around the Fermi energy) into the above scaling equation such that $J$ becomes dimensionless. After the transformation, the effective island-bath coupling term recovers a diagonal (but possibly highly nonlocal because of the transformation) form
\begin{align}\label{eq:HIB_diag}
    H_{IB} = \sum_{\alpha\ne\beta}2J_{\alpha\beta} \tilde{\gamma}_\alpha\tilde{\gamma}_\beta \tilde{c}_{{\beta}}^\dag \tilde{c}_{{\alpha}},
\end{align}
where $\tilde{\gamma}_\alpha = \sum_{\alpha'}{\gamma}_{\alpha'}R_{\alpha'\alpha}$ and $\tilde{c}_{{\alpha}} = \sum_{\alpha'}{c}_{\bm{r}_{\alpha'}}S_{\alpha'\alpha}$. It is worth noting that despite the apparent diagonal form of Eq.~\eqref{eq:HIB_diag}, the generic non-orthogonality of the matrix $S$ leaves the operators $\tilde{c}$ not satisfying the canonical fermionic anti-commutation relations, therefore extra care must be taken in interpreting of the effective coupling in Eq.~\eqref{eq:HIB_diag} physically.

\begin{figure}[t]
\centering
\includegraphics[width=\linewidth]{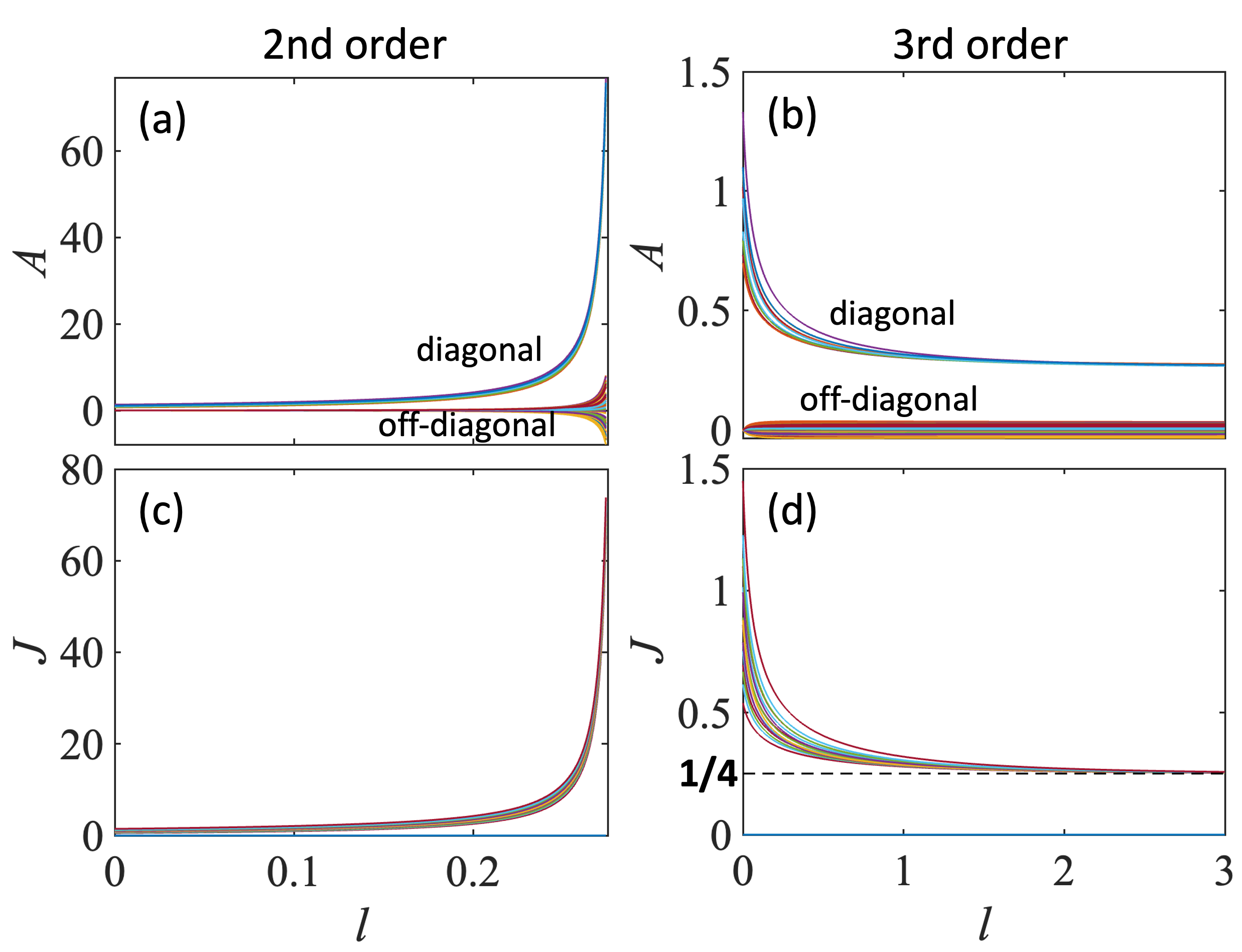}
\caption{Numerically solved renormalization group flows of (a, b) the original island-bath coupling coefficients $A_{\alpha\beta}^{\mu\nu}$ as in Eq.~\eqref{eq:RG_org}, and (c, d) the transformed diagonal entries $J_{\alpha\beta}$ as in Eq.~\eqref{eq:RG_J}, with the scaling equation including corrections up to (a, c) the second order and (b, d) the third order, respectively. Note that both $A$ and $J$ are made dimensionless by multiplying to $\rho_0$. We have used $M=6$ in this example. For generality, we have randomly determined each value of $\lambda$'s in the range $[0.8,1.2]$ and similarly the off-diagonal elements of $f$ in $[-0.2,0.2]$ (the diagonal elements of $f$ are set to $1$).  The RG flow behaviors illustrated here as well as the consistency between the solutions of Eqs.~\eqref{eq:RG_org} and \eqref{eq:RG_J} are completely general.}\label{fig:flows}
\end{figure}

Both the original scaling equation \eqref{eq:RG_org} and the reduced one \eqref{eq:RG_J} can be numerically solved and compared in generic settings of $\{\lambda_\alpha\}$ and $f$. We present the typical results of such numerical solutions, up to the second-order and the third-order corrections, respectively, in Fig.~\ref{fig:flows}. From the numerical results, we observe that the effective island-bath coupling always flows towards the strong and isotropic coupling limit at increasing lower energy: up to the second order correction, the non-zero elements of $J$ diverge at large $l$ (see Fig.~\ref{fig:flows}); up to the third order correction, the non-zero elements of $J$ converge to an identical value at large $l$ (see Fig.~\ref{fig:flows}). Indeed, this fixed point can be derived straightforwardly from Eq.~\eqref{eq:RG_J}. Particularly, by letting $J_{\alpha\ne\beta}=J_*$ and $J_{\alpha=\beta}=0$, the fixed-point condition for the scaling equation \eqref{eq:RG_J} becomes
\begin{align}\label{eq:fiexed_point_eq}
    (M-2)J_*^2 + 4J_*^3 - 4(M-1)J_*^3 = 0,
\end{align}
which clearly has a nontrivial, $M$-independent, solution $J_*=1/4$. This fixed point is precisely what we obtain from numerics with generic physical settings in our model, presented in Fig.~\ref{fig:flows}. Furthermore, this fixed point is a stable one as can be seen from the following perturbation analysis. Let $J_{\alpha\ne\beta} = 1/4 + {\delta J}_{\alpha\beta}$, where $\vert{\delta J}_{\alpha\beta}\vert\ll 1/4$. To linear order, Eq.~\eqref{eq:RG_J} becomes (see Appendix \ref{app:fixed_point})
\begin{equation}
    \frac{\dd{\delta J}_{\alpha\beta}}{\dd l} = -\frac{M-2}{4}{\delta J}_{\alpha\beta},
\end{equation}
which implies ${\delta J}_{\alpha\beta}(l) \propto \exp(-\frac{M-2}{4}l)$.
In other words, $J$ always flows back to the fixed point $J_*$ from a small deviation as long as $M>2$.

\section{Discussion and Outlook}\label{sec:discussion}

The construction and characterization of Majorana islands stands to be a major challenge in the development of topological quantum computation. In this work we investigate Majorana islands tunnelling-embedded in metallic bath, which would be a natural setup for a variety of Majorana platforms. We have insofar discussed three theoretical aspects of such embedded Majorana islands --- all these aspects emphasize the presence and the consequence of electronic interference in the metal, which is a unique feature connecting the nonlocality of Majorana qubit states with the Kondo-type physics. In this last section we discuss the limitation and the connection of these aspects, and lay out questions for follow-up research.

First, topological ground-state degeneracy is the defining character of topological qubits as well as the key ingredient for the emergence of Kondo-type physics. In Sec.~\ref{sec:topo_deg} we have shown that the topological degeneracy in a Majorana island persists even when it's coupled to a metallic bath. There is one caveat, however, regarding this conclusion. The caveat is technically associated with the assumption that the phase factor $\mathrm{e}^{\ii\phi/2}$ in Eq.~\eqref{eq:HT0}, being interpreted as an operator acting on the superconducting island state to add one electron, is independent on the Majorana index $\alpha$ or the position $\bm{r}$. If this assumption is relaxed, then the self-energy due to the metallic bath may indeed lift the energy degeneracy via the fluctuation of the paring phase \cite{li_spontaneous_2013} by $\delta E \propto \rho_0\lambda^2 \langle\sin\delta\phi\rangle /(\delta r)^{\frac{d-1}{2}}$. The validity of this assumption requires a more careful examination of the Majorana island in terms of a particle-conserving theory, which is a part of our on-going work and will be reported elsewhere.

Second, the mapping from the generic island-bath coupling term in Eq.~\eqref{eq:HIB} to a generalized Kondo coupling term derived in Sec.~\ref{sec:ham_eff} is entirely formal and requires the algebraic condition in Eq.~\eqref{eq:Acom} to be satisfied. The purpose of this mapping, as explained also in Sec.~\ref{sec:ham_eff}, is to take advantage of existing methods and conclusions in the Kondo model and its variants. For Majoranas that are in contact with individual leads, ${A}_{\alpha\beta}^{\mu\nu} \propto \delta_{\alpha\beta}^{\mu\nu}$ therefore Eq.~\eqref{eq:Acom} is naturally satisfied. For the embedded Majorana island considered in our case, however, the presence of electronic interference, represented by a non-diagonal matrix $f_{\mu\nu}$, implies that the preserved island-bath coupling form under the RG process is given by Eq.~\eqref{eq:HIB_diag} which contains a non-orthogonal transformation of operators ${c}_{\bm{r}}$ and hence does not satisfy Eq.~\eqref{eq:Acom} generically. This poses a new scenario that distinguishes our model with conventional generalization of Kondo models. To fully understand the consequence of this subtle but significant distinction is one major goal of our follow-up work.

Third and closely related to the second above, our scaling analysis in this work has been based on perturbations and hence is restricted in reaching a comprehensive conclusion or, more importantly, computing physical observables. To solve this problem properly requires an unbiased numerical approach such as numerical renormalization group and its adaptations \cite{bulla2008numerical}. By performing a full-scaled numerical calculation, we will be able to not only determine the RG flow more definitively, but also obtain experimentally measurable signatures such as real-space images of the Kondo cloud due to the embedded Majorana island and its topological degeneracy. Further extensions may also include electronic transport in connected multiple nanowires hosting MZMs with quantum interference effect.

To conclude, we have investigated the model of Majorana islands tunneling embedded in a simply-connected metallic bath and demonstrated its connection to conventional Kondo physics as well as the crucial distinction owing to the interplay between the Majorana-led nonlocality and the electronic interference. Our work serves as the necessary first step in tackling this experimentally relevant and pressing problem towards the detection of topological qubits.

\section{ACKNOWLEDGEMENTS}

We acknowledge the participation of Berthold J\"{a}ck and Ali Yazdani at the early stage of this work. We are grateful to  Zhi-Rui Yao, Chang-An Li, Kun Jiang, Xin Liu, Congjun Wu and Hong Ding for helpful discussions. Jian Li acknowledges the support from the National Natural Science Foundation of China under Project 92265201 and the Innovation Program for Quantum Science and Technology under Project 2021ZD0302704. Wen-Li Yang acknowledges the support from the National Natural Science Foundation of China (Grant No. 12247103) and the Shaanxi Fundamental Science Research Project for Mathematics and Physics (Grant No. 22JSZ005). Rui Wang acknowledges the support from the National Natural Science Foundation of China (Grant No. 12322402 and No. 12274206).

\appendix

\section{Derivation of Eq.~\eqref{eq:p_h_symmetry}}\label{app:phs}

To derive Eq.~\eqref{eq:p_h_symmetry}, let us start from the general definitions of the greater and the lesser correlation functions:
\begin{align}
  &C^{>}(A,B;t) \equiv -\ii \langle A(t) B(0)\rangle \nonumber \\
  &\quad= -\ii \Tr \left[ \mathrm{e}^{-\beta H}\mathrm{e}^{\ii t H}A \mathrm{e}^{-\ii t H} B \right] / \Tr \mathrm{e}^{-\beta H}, \\
  &C^{<}(A,B;t) \equiv -\ii \eta \langle B(0) A(t)\rangle \nonumber \\
  &\quad= -\ii\eta \Tr \left[ \mathrm{e}^{-\beta H} B \mathrm{e}^{\ii t H}A \mathrm{e}^{-\ii t H} \right] / \Tr \mathrm{e}^{-\beta H},
\end{align}
where $A$ and $B$ are two operators; $\eta=+1$ when they are bosonic and $\eta=-1$ when they are fermionic. It is easy to check that
\begin{align}
  C^{>}(A,B;t)^* &= -\eta C^{<}(A^\dag,B^\dag;t),  \\
  C^{<}(A,B;t)^* &= -\eta C^{>}(A^\dag,B^\dag;t).
\end{align}
The retarded correlation function is subsequently defined as
\begin{equation}
  C^{R}(A,B;t) = \theta(t) \left[C^{>}(A,B;t) - C^{<}(A,B;t)\right],
\end{equation}
which satisfies
\begin{align}
  &C^{R}(A,B;\omega)
  \equiv \int \dd t\, \mathrm{e}^{\ii\omega t}\, C^{R}(A,B;t)   \\
  &= \int \dd t\, \mathrm{e}^{\ii\omega t}\, \theta(t) \left[C^{>}(A,B;t) - C^{<}(A,B;t)\right]   \\
  &= \int \dd t\, \mathrm{e}^{\ii\omega t}\, \theta(t) (-\eta) \left[C^{<}(A^\dag,B^\dag;t)^* - C^{>}(A^\dag,B^\dag;t)^*\right] \\
  &= \eta\, C^{R}(A^\dag,B^\dag;-\omega)^*.
\end{align}
By setting $A=c_{\bm{r}}$, $B=c_{\bm{r}'}^\dag$ and $\eta = -1$, the above general relation becomes Eq.~\eqref{eq:p_h_symmetry}.

\section{Derivation of the scaling equation Eq.~\eqref{eq:RG_org}}\label{app:rgeq}

The strategy to derive the scaling equation is to make canonical transformations that gradually decouple the low energy sector (with momentum labeled by $k$ and energy $|\xi_k|<D-\delta D$) and the high energy sector (with momentum labeled by $p$ and energy $D-\delta D<|\xi_p|<D$) of the metallic bath electronic states up to increasingly higher order of the bare island-bath coupling coefficients.

To start with, the metallic bath can be written as
\begin{equation}\label{eq:H0}
  H_0 = \sum_{|\xi_k|<D-\delta D}\xi_k c_{k}^\dag c_{k} +
  \sum_{D-\delta D<|\xi_p|<D}\xi_p c_{p}^\dag c_{p},
\end{equation}
where $D$ is the cut-off energy and $\delta D$ is a small energy shell to be integrated out. The general effective coupling between Majorana island and the metallic bath is given by (cf. Eq.~\eqref{eq:HIB} in the main text)
\begin{eqnarray}
   H_{IB} &=& \sum_{\alpha,\beta,\mu,\nu} A_{\alpha\beta}^{\mu\nu} \gamma_{\alpha}\gamma_{\beta} c_{r_\nu}^\dag c_{r_\mu} \n
   &=& \sum_{\alpha,\beta,\mu,\nu} \sum_{q,q^\prime} A_{\alpha\beta}^{\mu\nu} \mathrm{e}^{\ii (q^\prime r_\mu - q r_\nu)} \gamma_{\alpha}\gamma_{\beta} c_q^\dag c_{q^\prime}  \n
   &=& \sum_{q,q^\prime} \mathcal{A}^{qq^\prime} c_q^\dag c_{q^\prime}\n
   &=& H'_{IB} + V,
\end{eqnarray}
where $\mathcal{A}^{qq^\prime}\equiv \sum_{\alpha,\beta,\mu,\nu} A_{\alpha\beta}^{\mu\nu} \mathrm{e}^{\ii (q^\prime r_\mu - q r_\nu)} \gamma_{\alpha}\gamma_{\beta} = (\mathcal{A}^{q'q})^\dag$, and
\begin{align}
  H'_{IB} &= \sum_{k,k^\prime} \mathcal{A}^{kk^\prime} c_k^\dag c_{k^\prime} + \sum_{p,p^\prime} \mathcal{A}^{pp^\prime} c_p^\dag c_{p^\prime}, \label{eq:HIBprime} \\
  V &= \sum_{k,p} \mathcal{A}^{kp} c_k^\dag c_{p} + \mathrm{h.c.}. \label{eq:V}
\end{align}

We seek to derive
\begin{align}
  h^{(2)} &= \frac{1}{2}[\Omega_1,V], \label{} \\
  h^{(3)} &= \frac{1}{2}[\Omega_2,V], \label{third_order_eff_H}
\end{align}
as the second-order and the third-order corrections, respectively, where $\Omega_1$ and $\Omega_2$ satisfy
\begin{equation}\label{eq:omega1}
  V + [\Omega_1,H_0] = 0,
\end{equation}
and
\begin{equation}\label{eq:omega2}
  [\Omega_2,H_0] + [\Omega_1,H'_{IB}] = 0,
\end{equation}
respectively.

Substituting Eqs.~\eqref{eq:H0} and \eqref{eq:V} into Eq.(\ref{eq:omega1}), we obtain
\begin{equation}\label{eq:Omega1}
  \Omega_1 = \sum_{k,p} C^{kp} c_k^\dag c_{p} - \mathrm{h.c.},
\end{equation}
with
\begin{equation}\label{}
  C^{qq^\prime} = \frac{\mathcal{A}^{qq^\prime}}{\xi_q-\xi_{q^\prime}} = -(C^{q'q})^\dag.
\end{equation}
This further leads to
\begin{small}
\begin{eqnarray}
  h^{(2)} &=&  \frac{1}{2}[\Omega_1,V] \n
  &=& \frac{1}{2} \sum_{k,k',p,p'} [C^{kp} c_k^\dag c_{p}, \mathcal{A}^{p^\prime k^\prime} c_{p^\prime}^\dag c_{k^\prime}] + \mathrm{h.c.}  \n
  &=& \sum_{k,k',p,p'} (C^{kp}\mathcal{A}^{p^\prime k^\prime}c_k^\dag c_{p}c_{p^\prime}^\dag c_{k^\prime}
  - \mathcal{A}^{p^\prime k^\prime} C^{kp} c_{p^\prime}^\dag c_{k^\prime} c_k^\dag c_{p})  \n
  &\approx& \frac{{\rho_0}\delta D}{-D}\sum f_{\mu\nu^\prime}A_{\alpha\beta}^{\mu\nu} A_{\alpha^\prime\beta^\prime}^{\mu^\prime\nu^\prime} [\gamma_{\alpha}\gamma_{\beta}, \gamma_{\alpha^\prime}\gamma_{\beta^\prime}] c_{r_{\nu}}^\dag c_{r_{\mu^\prime}} \n
   &=& \frac{2{\rho_0} \delta D}{-D}\sum (A_{\alpha\eta}^{\mu\nu}A_{\eta\beta}^{\mu^\prime\nu^\prime}- A_{\beta\eta}^{\mu\nu}A_{\eta\alpha}^{\mu^\prime\nu^\prime})f_{\mu\nu^\prime}\gamma_\alpha\gamma_\beta c_{r_{\nu}}^\dag c_{r_{\mu^\prime}} \n
   &=& \sum_{\alpha\beta\mu^\prime\nu} \frac{2{\rho_0} \delta D}{-D}(P_{\alpha\beta}^{\nu\mu^\prime}-P_{\beta\alpha}^{\nu\mu^\prime})\gamma_\alpha\gamma_\beta c_{r_{\nu}}^\dag c_{r_{\mu^\prime}},
\end{eqnarray}
\end{small}where the summation runs over all indices. 
Since the Fermi energy $\varepsilon_F$ is much larger than the energy scale of the 
RG process, we replace the magnitude of $k$ and $p$ by $k_F=\frac{\varepsilon_F}{2m}$.
Here we assume that $\xi_k \ll \xi_p = \pm D$ and 
the following identity is used
\begin{small}
\begin{equation}\label{}
  \sum A_{\alpha\beta}^{\mu\nu} A_{\alpha^\prime\beta^\prime}^{\mu^\prime\nu^\prime}[\gamma_{\alpha}\gamma_{\beta}, \gamma_{\alpha^\prime}\gamma_{\beta^\prime}] = \sum 2(A_{\alpha\eta}^{\mu\nu}A_{\eta\beta}^{\mu^\prime\nu^\prime}- A_{\beta\eta}^{\mu\nu}A_{\eta\alpha}^{\mu^\prime\nu^\prime})\gamma_\alpha \gamma_\beta;
\end{equation}
\end{small} 
we denote
\begin{equation}\label{}
  P_{\alpha\beta}^{\nu\mu^\prime} \equiv \sum_{\eta\mu\nu^\prime}A_{\alpha\eta}^{\mu\nu}A_{\eta\beta}^{\mu^\prime\nu^\prime}f_{\mu\nu^\prime},
\end{equation}
for convenience of expression; $f_{\mu\nu} = f_{\nu\mu}$ is the form factor defined as $f_{\mu\nu} \equiv \sum_{p=k_F} \mathrm{e}^{\ii \bm{p}\cdot (\bm{r}_{\mu}-\bm{r}_{\nu})}/\sum_{p=k_F} 1$, which physically represents the electron propagator in the metal. Explicitly, $f$ is an angular integral at the Fermi surface given by 
\begin{equation}
    f_{\mu\nu} =
    \begin{cases}
        J_0(k_F |\bm{r}_\mu - \bm{r}_\nu|) & \mathrm{for \ 2D \ bath}, \\
        \mathrm{sinc}{(k_F |\bm{r}_\mu - \bm{r}_\nu|)}  & \mathrm{for \ 3D \ bath}.
    \end{cases}
\end{equation}

Substituting ~\eqref{eq:HIBprime} and ~\eqref{eq:Omega1} into Eq.(\ref{eq:omega2}), we can obtain
\begin{eqnarray}
  \Omega_2 &\approx& \frac{1}{(\xi_k-\xi_p)(\xi_k-\xi_p + \xi_{k_1}-\xi_{k_1^\prime})} A^{kp} A^{k_1 k_1^\prime} \n &&[\gamma_{\alpha}\gamma_{\beta} c_{k}^\dag c_{p}, \gamma_{\alpha_1}\gamma_{\beta_1} c_{k_1}^\dag c_{k_1^\prime}] + \mathrm{h.c.},  \n
\end{eqnarray}
here we have dropped the $p,p^\prime$ term in ~\eqref{eq:HIBprime} under the assumption that only the linear order in $\delta D/D$ contributions to the RG correction are considered.
Then, the third order correction is given by
\begin{widetext}
\begin{small}
\begin{eqnarray}
  h^{(3)} &=& \frac{1}{2}[\Omega_2, V] \n
  &=& \sum\frac{A_{\alpha\beta}^{\mu\nu} A_{\alpha_1\beta_1}^{\mu_1\nu_1} A_{\alpha_2\beta_2}^{\mu_2\nu_2} \mathrm{e}^{-\ii k r_\nu + \ii p r_\mu - \ii k_1 r_{\nu_1} + \ii k_1^\prime r_{\mu_1} -\ii p_2 r_{\nu_2} + \ii k_2 r_{\mu_2} } }{2(\xi_k-\xi_p)(\xi_k-\xi_p + \xi_{k_1}-\xi_{k_1^\prime})} \times[[\gamma_{\alpha}\gamma_{\beta} c_{k}^\dag c_{p}, \gamma_{\alpha_1}\gamma_{\beta_1} c_{k_1}^\dag c_{k_1^\prime}],\gamma_{\alpha_2}\gamma_{\beta_2} c_{p_2}^\dag c_{k_2}] + \mathrm{h.c.} \n
  &\approx& \sum \frac{{\rho_0}^2\delta D}{2D}  f_{\nu\mu_2}  f_{\nu_2\mu}  A_{\alpha\beta}^{\mu\nu} A_{\alpha_1\beta_1}^{\mu_1\nu_1} A_{\alpha_2\beta_2}^{\mu_2\nu_2} [[\gamma_{\alpha}\gamma_{\beta}, \gamma_{\alpha_1}\gamma_{\beta_1}],\gamma_{\alpha_2}\gamma_{\beta_2}] c_{r_{\nu_1}}^\dag c_{r_{\mu_1}}  \n
  &=& \sum \frac{2{\rho_0}^2\delta D}{D} f_{\nu\mu_2}f_{\nu_2\mu} (A_{\alpha\eta}^{\mu\nu} A_{\eta\beta}^{\mu_1\nu_1}-A_{\beta\eta}^{\mu\nu} A_{\eta\alpha}^{\mu_1\nu_1})  \times A_{\alpha_2\beta_2}^{\mu_2\nu_2} [\gamma_{\alpha}\gamma_{\beta},\gamma_{\alpha_2}\gamma_{\beta_2}] c_{r_{\nu_1}}^\dag c_{r_{\mu_1}} \n
  &=& \sum \frac{2{\rho_0}^2\delta D}{D} (Q_{\alpha\beta}^{\mu_1\nu_1}-Q_{\beta\alpha}^{\mu_1\nu_1})\gamma_\alpha\gamma_\beta c_{r_{\nu_1}}^\dag c_{r_{\mu_1}},
\end{eqnarray}
\end{small}
\end{widetext}where the summation runs over all indices, and denote
\begin{small}
\begin{equation}\label{}
  Q_{\alpha\beta}^{\mu_1\nu_1}\equiv \sum_{\lambda\eta\mu\nu\mu_2\nu_2} f_{\nu\mu_2}f_{\nu_2\mu} (A_{\alpha\eta}^{\mu\nu}A_{\eta\lambda}^{\mu_1\nu_1} A_{\lambda\beta}^{\mu_2\nu_2}
  - A_{\lambda\eta}^{\mu\nu} A_{\eta\alpha}^{\mu_1\nu_1} A_{\lambda\beta}^{\mu_2\nu_2}).
\end{equation}
\end{small}Then the RG equation is given by Eq.\eqref{eq:RG_org}. 

This $f$ dependent RG equation can always be transformed into $f$ independent one, Eq.~\eqref{eq:RG_tfm}, through the transformation Eq.\eqref{eq:tfm_A2A_tilde}. 
By inserting a physical initial point $\tilde{A}_{\alpha\beta}^{\mu\nu}(0) = J_{\alpha\beta}(0)\delta_{\alpha\beta}^{\mu\nu}$ (where the diagonal elements $J_{\alpha\alpha}$ are zero) into Eq.~\eqref{eq:RG_tfm}, we obtain
\begin{small}
    \begin{eqnarray}
    \frac{\dd J_{\alpha\beta}}{\dd l} \delta_{\alpha\beta}^{\mu\nu} &=& \delta^{\mu\nu}_{\alpha\beta}\{(J^2)_{\alpha\beta} + 4(J_{\alpha\beta})^3 - 2J_{\alpha\beta}[(J^2)_{\alpha\alpha}+(J^2)_{\beta\beta}]\}. \n
\end{eqnarray}
\end{small}The RHS is equal to a symmetric matrix multiplied by $\delta^{\mu\nu}_{\alpha\beta}$, which implies that the diagonal form of $\tilde{A}_{\alpha\beta}^{\mu\nu}(l)$ for all $l$ is preserved under the RG transformation. Consequently, the scaling equation
can be further reduced to a much simpler form Eq.~\eqref{eq:RG_J}. 

\section{Fixed point solutions}\label{app:fixed_point}
It is easy to check that $J_0$ with the matrix elements $J_{\alpha\neq\beta}=\sigma^\alpha\sigma^\beta/4$ (where $\sigma^\alpha = \pm 1$), $J_{\alpha=\beta}=0$  are fixed points of Eq.~\eqref{eq:RG_J}. Without loss of generality, we choose $\sigma^\alpha = 1$ for simplicity such that $J_{\alpha\neq\beta}=J_* = 1/4$.
We now illustrate the stability of these fixed points from a perturbation perspective. 
For any small perturbations around these fixed points, namely, 
$J = J_0 + \delta J$ (the matrix elements are expressed as $J_{\alpha\neq\beta} = J_* + \delta J_{\alpha\neq\beta}$, $J_{\alpha=\beta}=0$, where $\vert\delta J_{\alpha\beta}\vert\ll 1/4$),
to linear order, the first term in the RHS of the RG eqation Eq.~\eqref{eq:RG_J} is given by
\begin{equation}
    J^2 = (J_0+\delta J)^2 \approx J_0^2 + J_* \delta\tilde{J},
\end{equation}
where $J_* \delta\tilde{J} \equiv J_0 \delta J + \delta J J_0$, which implies that
\begin{eqnarray}\label{eq:dJ_tilde}
    \delta\tilde{J}_{\alpha\beta} &=& \sum_{\eta} \tilde{I}_{\alpha\eta} \delta J_{\eta\beta} + \delta J_{\alpha\eta} \tilde{I}_{\eta\beta} \n
    &=& \sum_{\eta\neq\alpha,\beta} \delta J_{\eta\beta} + \delta J_{\alpha \eta},
\end{eqnarray}
is defined for calculation convenience. Here
$\tilde{I}_{\alpha\neq \beta}=1$, $\tilde{I}_{\alpha\alpha}=0$.
The second and third terms in the RHS of the RG equation Eq.~\eqref{eq:RG_J} are given by 
\begin{equation}
    4(J_{\alpha\beta})^3 = 4(J_*+\delta J_{\alpha\beta})^3 \approx 4(J_*^3+3J_*^2\delta J_{\alpha\beta}),
\end{equation}
and 
\begin{widetext}
    \begin{eqnarray}
    & &-2J_{\alpha\beta}\left( \left(J^2\right)_{\alpha\alpha} +\left(J^2\right)_{\beta\beta} \right) \n
    &\approx& -2 (J_*+\delta J_{\alpha\beta}) \left( \left(J_0^2\right)_{\alpha\alpha} +\left(J_0^2\right)_{\beta\beta} + J_*\delta \tilde{J}_{\alpha\alpha} + J_* \delta \tilde{J}_{\beta\beta} \right)  \n
    &\approx& -2\left( J_* \left(J_0^2\right)_{\alpha\alpha} +J_* \left(J_0^2\right)_{\beta\beta} +  2(M-1)J_*^2 \delta J_{\alpha\beta}
    + J_*^2 \left( \delta \tilde{J}_{\alpha\alpha} + \tilde{J}_{\beta\beta} \right)  \right),
    \end{eqnarray}
\end{widetext}
respectively, where the identity Eq.~\eqref{eq:dJ_tilde} has been used. In the process of RG transformation, the RG equation for the disturbance $\delta J_{\alpha\beta}$ near the fixed point is given by
\begin{widetext}
    \begin{eqnarray}
    \frac{\dd \delta J_{\alpha\beta}}{\dd l} &=& J_* \delta\tilde{J}_{\alpha\beta} + 12 J_*^2 \delta J_{\alpha\beta} - 4(M-1)J_*^2 \delta J_{\alpha\beta} - J_*^2 \left( \delta\tilde{J}_{\alpha\alpha} + \delta \tilde{J}_{\beta\beta} \right) \n
    &=& -\frac{M-2}{4} \delta J_{\alpha\beta},
    \end{eqnarray}
\end{widetext}
with the solution
\begin{equation}
	\delta J_{\alpha\beta}(l) \propto \exp\left(-\frac{M-2}{4}l\right).
\end{equation}
The magnitude of small perturbations around a specific fixed point of this type can then be seen to flow to zero.

For each $M$, we can also use fixed points of lower $2 < M^\prime < M$ by setting some indices to zero. If we only have two non-zero indices in $J_{\alpha\beta}$, Eq.~\eqref{eq:fiexed_point_eq} tells us that it is a fixed point regardless of the magnitude of $J_0$. This gives us fixed lines. We can not only have one non-zero index, since this would correspond to a diagonal $J_{\alpha\beta}$, which is equivalent to a trivial zero solution.
Moreover, if we consider block diagonal solutions, Eq.~\eqref{eq:RG_J} decouples into separate equations for each respective block which we can solve in the same way as done above. It is worth noting that the fixed point does not depend on the effective number of MFs $M^\prime$ ($M^\prime \neq 2$) and the geometric distribution of MFs.

\bibliography{ref/Jinxing_ref}

\end{document}